\documentclass[aps, prb, reprint, lengthcheck, footinbib, showpacs, superscriptaddress, citeautoscript]{revtex4-1}

\usepackage{amsmath}
\usepackage{amssymb}
\usepackage{braket}
\usepackage{graphicx}
\usepackage{euscript}
\usepackage[cmyk]{xcolor}
\usepackage[unicode, colorlinks, urlcolor=blue, citecolor=blue, linkcolor=blue]{hyperref}
\usepackage[all]{hypcap}
\usepackage[utf8]{inputenc}
\usepackage[english]{babel}
\usepackage{cmap}
\usepackage{blindtext}

\begin{document}

%\def\affiSOLAB{Spin Optics Laboratory, Saint~Petersburg State University, 198504 St.~Petersburg, Russia}

%\large

\title{Decrease of heavy-hole exciton mass induced by uniaxial stress in GaAs/AlGaAs quantum well}

\author{D.~K.~Loginov}
\affiliation{Spin Optics Laboratory, St. Petersburg State University, 198504 St.~Petersburg, Russia}
\author{P.~S.~Grigoryev}
\affiliation{Spin Optics Laboratory, St. Petersburg State University, 198504 St.~Petersburg, Russia}
\author{E.~V.~Ubiyvovk}
\affiliation{Department of Physics, St. Petersburg State University, 198504 St.~Petersburg, Russia}
\author{Yu.~P.~Efimov}
\affiliation{Resource Center ``Nanophotonics'', St. Petersburg State University, 198504 St.~Petersburg, Russia}
\author{S.~A.~Eliseev}
\affiliation{Resource Center ``Nanophotonics'', St. Petersburg State University, 198504 St.~Petersburg, Russia}
\author{V. A. Lovtcius}
\affiliation{Resource Center ``Nanophotonics'', St. Petersburg State University, 198504 St.~Petersburg, Russia}
\author{Yu.~P.~Petrov}
\affiliation{Resource Center ``Nanophotonics'',  St. Petersburg State University, 198504 St.~Petersburg, Russia}
\author{I.~V.~Ignatiev}
\affiliation{Spin Optics Laboratory,  St. Petersburg State University, 198504 St.~Petersburg, Russia}
%\maketitle

\date{\today }

\begin{abstract}
It is  experimentally shown that the pressure applied along the twofold symmetry axis of a heterostructure with a wide GaAs/AlGaAs quantum well leads to considerable modification of  the polariton reflectance spectra. This effect is treated as the stress-induced decrease of the heavy-hole exciton mass. Theoretical modeling of the effect supports this assumption. The 5\%-decrease of the exciton mass is obtained at pressure $P=0.23$ GPa. 
\end{abstract}

\pacs{71.35.-y, 71.36.+c, 71.70.Fk, 78.67.De}

\maketitle

\section*{Introduction}

In a number of structures like topological insulators~\cite{Qi,HgTe}, superconductors~\cite{Pashkin, Matsunaga} and graphen \cite{Novoselov2, ScientReport}, the dispersion of the electron and hole energy near the $\Gamma$-point has a form of the Dirac cone that corresponds to the massless charged carriers.
The time-reversal or spontaneous spatial symmetry breaking is capable, however, to make the charge carriers in these semiconductors massive (see, e.g., Refs.~\onlinecite{Qi, Pashkin, ScientReport}). The generated carrier mass is described by the nondiagonal terms of the Dirac Hamiltonian which are  independent of wave vector $\mathbf k$ of the carrier.
In Ref.~\onlinecite{Qi}, this mass is called as ``Dirac'' in order to distinguish it from the mass described by the matrix elements depending on $\mathbf k^2$.
The latter is called as ``Newtonian'' in the cited work.

Dirac mechanisms of the mass generation have been mainly discussed for the two-dimensional and one-dimensional semiconductors with the dominant linear over $\mathbf k$ terms in the carrier Hamiltonian.
At the same time, in the bulk semiconductors and heterostructures where the carriers are described by a Hamiltonian with the dominant quadratic over $\mathbf k$ terms, 
the generation of additions to the effective mass is also possible due to mechanisms similar to the Dirac ones.
For example, in the topological crystalline insulators, the carrier energy quadratically depends on wave vector $\mathbf k$ though the band gap is close to zero~\cite{Liang, Sun}. Any small change of crystal symmetry gives rise to the appearance of an energy gap and, as a consequence, to the change of carrier effective masses.

 In a bulk GaAs crystal, the exciton energy dispersion law is quadratic in wave vector too. When an uniaxial stress is applied along axis [100] to the GaAs crystal, the change (convergence) of masses of the heavy- and light-hole excitons should be observed as it was theoretically predicted in Ref.~\onlinecite{LogTrifIg2014}. %
A formal description of this effect is similar to the description of the Dirac mechanism of the carrier mass generation in topological insulators and superconductors because it is described by the nondiagonal Hamiltonian matrix elements independent of $\mathbf k$.

Present work is devoted to the theoretical and experimental studies of the mass convergence effect in polariton reflectance spectra of a wide quantum well (QW) in the presence of a uniaxial stress applied along [110] axis.
From the experimental point of view, this direction is convenient because it is perpendicular to the cleavage plane of the sample, which coincides with the crystal plane (110).
Therefore, the pressure along this direction prevents unwanted breakup of the sample.
Theoretical consideration of the pressure effect is similar to that performed in Ref.~\onlinecite{LogTrifIg2014} for the pressure applied along the [100] axis.

\section{Exciton Hamiltonian in the stress field}

\label{hamiltonian}

We consider an exciton in a crystal with the zinc blende
symmetry propagating along the $Z$ axis axis ([001] crystal axis).
This direction will be regarded as the axis of the carrier's angular momentum quantization.

Exciton states observed in optical experiments with the GaAs semiconductors are formed by the states of the twofold-degenerate conduction band $\Gamma_6$ and fourfold-degenerate valence band $\Gamma_8$.
Wave functions of electrons and holes in this approximation are represented by the two-component and four-component plane waves~\cite{EOKane}, respectively.

In the present work, we neglect the effects related to the corrugation of valence band~\cite{EOKane} since they are small and do not directly affect the discussed phenomenon.
By the same reason we ignore small effects of the exchange coupling of exciton states, of the nonparabolicity of exciton dispersion and of the coupling of the ground and excited exciton states~\cite{Zimmermann2000}.

The energy of an exciton moving in an unstrained crystal is~\cite{EOKane}:
\begin{equation}
H^{(0)}_{Xh,l}=E_X+\frac{\hbar^2 K^2}{2M_{h,l}},
\label{eq:zeroenergy}
\end{equation}
where $E_X=E_g-R_X$ is the energy of the exciton ground state. 
Here $E_g$ is the band gap and $R_X$ is the exciton binding energy (Rydberg),  $K=K_z$ is the exciton wave vector. 
Hereafter we assume that $K_x = K_y = 0$ that is the exciton propagates along the $z$-axis. 
The 
translational masses of the heavy-hole (denoted by index ``$h$'') and light-hole (denoted by index ``$l$'') excitons are defined by expressions:
$$
M_{h}=m_e+m_h,\text{ and } M_{l}=m_e+m_l.
$$  
Here $m_{e}$ is the electron effective mass, $m_{h,l}=m_0/(\gamma_1\mp2\gamma_2)$ is the effective masses of a heavy (sign ``-'') and light (sign ``+'') holes, $\gamma_1$ and $\gamma_2$ are the Luttinger parameters, $m_0$ is a mass of a free electron.

In the framework of approximations introduced above, the Hamiltonian of the exciton moving in the unstrained crystal, according to Ref.~\onlinecite{EOKane}, may be written as the $8\times8$ matrix, consisting of two identical diagonal blocks of the size $4\times4$, which nonzero elements has form~(\ref{eq:zeroenergy}).
The first of these blocks describes optically active exciton states with projections of the exciton spin $J_z=\pm1$, whereas the second one represents optically inactive states with projections $\pm2$ and $0$ for heavy and light excitons, respectively.

Basis wave functions of such Hamiltonian taking into account the spin moment of exciton are written as:
\begin{equation}
|j,s\rangle_K=\nu_{j,s}\Psi(K),
\label{eq:wavepure}
\end{equation}
where $\nu_{j,s}$ is the eight-component spinor with one non-zero component equal unity; $j=\pm3/2,~\pm1/2$ and $s=\pm1/2$ are the projections of the spin moments of the hole and the electron, respectively, to the selected axis; $\Psi (K)$ is the coordinate part of the wave function.

Let us now consider the impact of the uniaxial stress $P$ applied along axis [110] on the exciton states.
Components of the strain tensor are described by following expressions~\cite{BirPikus}:
\begin{equation}
\begin{split}
&\varepsilon_{xx}=\varepsilon_{yy}=-(S_{11}+S_{12})\frac12P,~ \varepsilon_{zz}=-S_{12}P,\\
&\varepsilon_{xy}=-S_{44}\frac12P,~
\varepsilon_{xz}=\varepsilon_{yz}=0,
\end{split}
\label{eq:epsilonP}
\end{equation}
where $S_{11},~S_{12},~S_{44}$ are components of the elastic compliance tensor.
Such strain, particularly, leads to the effect, which is described by the linear over the wave vector additive terms in the electron-hole Hamiltonians~\cite{PikusMaruschak, LewWillatzen}.
In the case under consideration, these additive terms lead to coupling of the exciton $1s$- and $np$-states and, as a consequence, to the change of the exciton binding energy.
Nevertheless, this change is small comparing to the effects discussed hereafter.
Therefore, we do not consider it in present work.

The most remarkable effect of crystal strain is the alteration of valence band $\Gamma_8$,
which is described by the Bir-Pikus Hamiltonian~\cite{BirPikus, LewWillatzen}.
This Hamiltonian does not couple optically active states, $|j,s\rangle=|\pm3/2,\mp1/2\rangle,~|\pm1/2,\pm1/2\rangle$, with optically inactive states, $|j,s\rangle=|\pm3/2,\pm1/2\rangle,~|\pm1/2,\mp1/2\rangle$.
To simplify our analysis, we turn from exciton basis~(\ref{eq:wavepure}) to a new representation using following formulas:
\begin{equation}
\begin{split}
&|h\alpha\rangle=\frac{1}{\sqrt{2}}|3/2,-1/2\rangle \pm\frac{i}{\sqrt{2}}|-3/2,1/2\rangle,\\
&|l\alpha\rangle=\frac{1}{\sqrt{2}}|1/2,1/2\rangle \pm\frac{i}{\sqrt{2}}|-1/2,-1/2\rangle.
\end{split}
\label{eq:newbasis}
\end{equation}
Here the upper and lower signs correspond to indexes $\alpha=x'$ and $\alpha=y'$, respectively.
These states can be excited by the light polarized along axes $[110]$ and $[1\bar10]$, which are the axes of optical anisotropy in the presence of the uniaxial stress.

A matrix of the total exciton Hamiltonian constructed using these wave functions is composed of two blocks:
\begin{equation}
\hat H_{X}=\begin{pmatrix}
H_{h\alpha} & V\\
V^* & H_{l\alpha}\\
\end{pmatrix},\text{ where } \alpha=x',~y'.
\label{eq:hamexfull'}
\end{equation}
Here
\begin{equation*}
\begin{split}
& H_{hx'}=H_{hy'}=H^{(0)}_{Xh}-a(S_{11}+S_{12}) P+ b(S_{11}+S_{12})\frac P2,\\
&H_{lx'}=H_{ly'}=H^{(0)}_{Xl}-a(S_{11}+S_{12}) P- b(S_{11}+S_{12})\frac P2,\\
&V=idS_{44}\frac P4,
\end{split}
\end{equation*}
where $a$, $b$, and $d$ are the strain potentials~\cite{BirPikus}.
Since the states corresponding to $\alpha=x'$ and $\alpha=y'$ are decoupled, the problem for each such block can be solved independently.

We should note here that the uniaxial stress also leads to a coupling of the light-hole states and states of the split-off valence band, which is resulted in an alteration of the light-hole mass~\cite{BirPikus}.
We have found, however, that respective energy shifts of the exciton states caused by this coupling are small in comparison to those due to the heavy-hole and light-hole coupling. Therefore, we do not consider this effect in what follows.

The nondiagonal elements $V$ in Hamiltonian~(\ref{eq:hamexfull'}) describes the effect of the mass convergence stipulated by coupling of the heavy-hole and light-hole excitons.
The effect is observed in optical spectra of a heterostructure with the wide QWs.
To model these spectra one needs to take into account the impact of the mass convergence effect on the dielectric properties of medium.

\section{Permittivity in the presence of uniaxial stress}
\label{polariton}

For analysis of the reflectance spectra with taking the pressure-induced effects into account, we employ a model of the polariton wave interference in a wide QW described, e.g., in Refs.~\onlinecite{Kavokin, Markov, Loginov}.
It is supposed that the incident light is directed along the normal to the sample surface and linearly polarized along one of the optical anisotropy axes.

The exciton-photon interaction is described by the perturbation operator (see, e.g., Ref.~\onlinecite{ NozueCho}):
\begin{equation}
V_d =- (\mathrm d_h+\mathrm d_l)E^{(\alpha)},~(\alpha=x',~y'),
\label{eq:photex}
\end{equation}
where $E^{(\alpha)}$ is the electric field of a light wave and $d_{h,l}$ is the matrix elements of a dipole moment for the heavy-hole and light-hole exciton.
The square of the matrix elements can be expressed as: $\mathrm d^2_h=3\mathrm d^2_l=\hbar \omega_{LT}\epsilon_0\Omega$, where  $\hbar\omega_{LT}$ is the energy of the longitudinal-transverse splitting~\cite{Invchenko}, $\epsilon_0$ is the background permittivity and $\Omega$ is crystal volume. 

The wave function of a polariton is a linear combination of the ``pure'' exciton wave functions (\ref{eq:newbasis}):
\begin{equation}
\Psi_\alpha(K)=C_{vac}|vac\rangle+\sum_{\beta=h,l}C_{\beta\alpha}|\beta\alpha\rangle, 
\label{eq:wavemixex1}
\end{equation}
where $C_{\beta\alpha},~C_{vac}$ are the expansion coefficients and $|vac\rangle$ is the crystal vacuum state.

Dispersion relations, wave functions and permittivity in the case of exciton-photon interaction can be obtained using the method proposed in Refs.~\onlinecite{ LogTrifIg2014, NozueCho}.
To this end, one should first solve a system of equations for energies of the heavy-hole and light-hole excitons taking into account  interaction~(\ref{eq:photex}):
\begin{widetext}
\begin{equation}
(\mathbb H-\mathbb I\hbar\omega)\cdot\mathbb C=
\begin{pmatrix}
H_{h\alpha}-\hbar\omega & V & -\mathrm d_hE^{(\alpha)} \\
V^* & H_{l\alpha}-\hbar\omega & -\mathrm d_lE^{(\alpha)} \\
-\mathrm d_hE^{(\alpha)} & -\mathrm d_lE^{(\alpha)} & H_{vac}-\hbar\omega
\end{pmatrix} 
\cdot\begin{pmatrix}
C_{h\alpha}\\
C_{l\alpha} \\
C_{vac} \\
\end{pmatrix}=0.
\label{eq:dispdet}
\end{equation}
\end{widetext}
Here $H_{vac}=0$ is the energy of the vacuum state and $\hbar\omega$ is the photon energy.
The expansion coefficients $C_{\alpha}$ are easily found from the first two lines of this matrix equation if one assumes that the optical excitation is very weak, therefore, $C_{vac}\approx 1$.

The exciton-photon interaction leads also to a resonant polarization of medium, which depends on the exciton wave vector. This polarization is described by the expression~\cite{Kavokin, NozueCho}:
\begin{equation}
\begin{split}
&4\pi{\cal P}^{(\alpha)}=\frac1\Omega\langle \Psi_\alpha|e\cdot r| \Psi_\alpha\rangle=\frac1\Omega ({\mathrm d}_hC_{h\alpha}+{\mathrm d}_lC_{l\alpha}) \\
&=4\pi\chi_{l\alpha}E^{(\alpha)}+
4\pi\chi_{h\alpha}E^{(\alpha)}.
 \\
\end{split}
\label{eq:pertexmed}
\end{equation}
Here $4\pi\chi_{h\alpha}$ and $4\pi\chi_{l\alpha}$ are contributions to the optical susceptibility at the heavy-hole and light-hole exciton resonances, respectively. 
From formulae (\ref{eq:dispdet}) and (\ref{eq:pertexmed}) one can come to relations for $\chi_{h,l\alpha}$:
\begin{equation}
\begin{split}
& 4\pi\chi_{h\alpha}=\frac{\tilde H_{l\alpha}\hbar\omega_{LT}}{\tilde H_{h\alpha}\tilde H_{l\alpha}-|V|^2} \pm\frac1{\sqrt3}\frac{V\hbar\omega_{LT}}{\tilde H_{l\alpha}\tilde H_{h\alpha}-|V|^2},\\
&4\pi\chi_{l\alpha}=\frac13\frac{\tilde H_{h\alpha}\hbar\omega_{LT}}{\tilde H_{h\alpha}\tilde H_{l\alpha}-|V|^2}\pm\frac1{\sqrt3}\frac{V\hbar\omega_{LT}}{\tilde H_{l\alpha}\tilde H_{h\alpha}-|V|^2}.
\end{split}
\label{eq:Pol/pm/pm}
\end{equation}
Here 
$$\tilde H_{h,l\alpha}\equiv H_{h,l\alpha}-\hbar\omega+i\Gamma_{h,l},$$
where $\Gamma_{h,l}$ is a phenomenological parameter introduced to describe processes of energy dissipation. The upper and lower signs in relation~(\ref{eq:Pol/pm/pm}) correspond to the light polarization along the $x'$ and $y'$ axes. 

One should also keep in mind that, even in the absence of  exciton resonances, the uniaxial stress leads to an optical anisotropy of crystal due to the piezo-optic effect~\cite{etc1, etc2}.
This effect leads to an additional background permittivity, which, in the basis of light waves polarized along $x'$ and $y'$ axes, is described as~\cite{etc1, etc2}:
\begin{eqnarray}
-\delta\epsilon_{x'}=\delta\epsilon_{y'}=\pi_{44}P\equiv\delta\epsilon,
\label{eq:eo^p}
\end{eqnarray}
where $\pi_{44}$ is the component of the piezo-optic tensor.

The total permittivity of medium with account of contributions~(\ref{eq:Pol/pm/pm}) and~(\ref{eq:eo^p}) has the form:
\begin{equation}
\epsilon_{\alpha}(\omega,K)=\epsilon_0\pm\delta\epsilon+4\pi\chi_{h\alpha}+4\pi\chi_{l\alpha}, ~(\alpha=x',~y').
\label{eq:circulmatrel}
\end{equation}

In order to obtain dispersion relations for polariton eigenmodes, one should solve the following dispersion equation~\cite{Onzager}:
\begin{equation}
\epsilon_\alpha(\omega,K)=\frac{c^2K^2}{\omega^2},
\label{eq:disequat}
\end{equation}
where $c$ is the light velocity and $\epsilon_\alpha(\omega,K)$ is described by expression (\ref{eq:circulmatrel}).
Equation~(\ref{eq:disequat}) has independent solutions for $\alpha=x'$ and $\alpha=y'$, which correspond to two linear polarizations of the incident light.
For each polarization, the equation is a polynomial of the third order over $K^2$.
Its solutions are the eigenmodes, which differ from each other by the primary contribution of either the light component (photon-like, $p$-type dispersion branch) or states of the heavy-hole or light-hole exciton ($h$- or $l$-type exciton branch, respectively).
\begin{figure}[t] 
\includegraphics[clip,width=1.0\columnwidth]{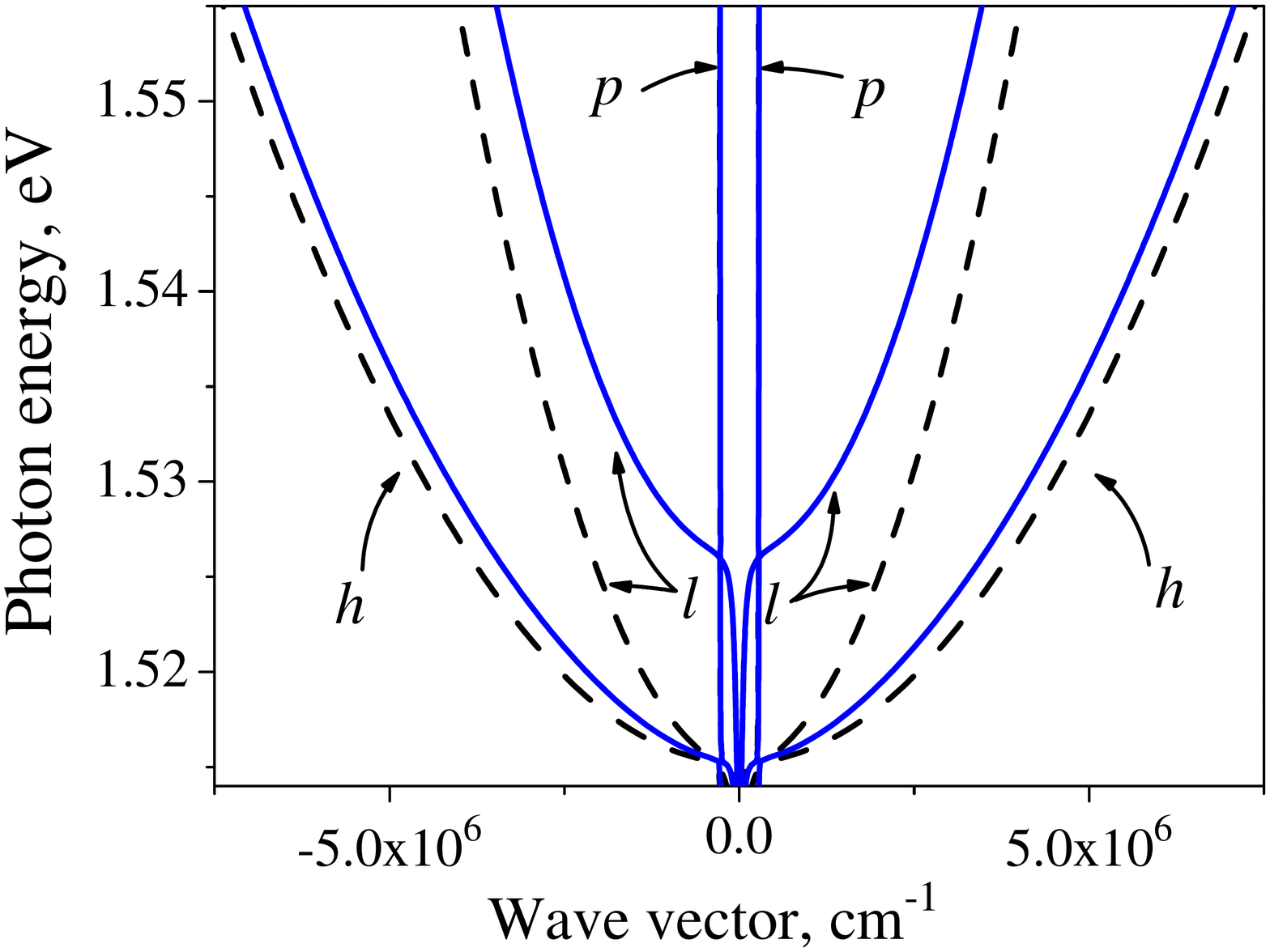}
\caption{Modification of the dispersion branches for  polaritons of $h$- and $l$-types in the bulk GaAs under the application of pressure along the [110] axis.  
The black dashed and blue solid curves are the dispersion branches at $P=0$ and $P=0.23$~GPa, respectively.
Notations $h$, $l$ and $p$ are explained in the text.}
\label{fig:Disp}
\end{figure}

Figure~\ref{fig:Disp} shows the dispersion curves for polariton eigenmodes calculated for strained and unstrained GaAs crystal.
The following material parameters were used:
$\hbar\omega_{LT}=0.08$ meV~\cite{Schultheis}, $\epsilon_0=12.56$~\cite{Stillman}, $m_e=0.067~m_0$~\cite{Bornstein}, $\gamma_1=6.8$, $\gamma_2=2.3$~\cite{Bornstein}, $E_g=1520$ meV~\cite{Bornstein}, $S_{11}=1.172\cdot 10^{-12}~\mathrm{cm^2/dyn}$, $S_{12}=-0.365\cdot 10^{-12}~\mathrm{cm^2/dyn}$, $S_{44}=1.68\cdot 10^{-12}~\mathrm{cm^2/dyn}$~\cite{Averkiev}, $\pi_{44}=1.5\text{ GPa}^{-1}$~\cite{etc1, etc2} and $d=-4.55$~eV, $a=-6.7$~eV, $b=-1.7$~eV~\cite{IvchenkoPikus}.

The deformation leads to reduction of the GaAs crystal symmetry from $T_d$ to $D_{2d}$.
Therefore, the dispersion curves of the exciton-like modes of $h$- and $l$-types are shifted to higher energies due to the diagonal terms of  Hamiltonian~(\ref{eq:dispdet}).
The $l$-type mode is shifted stronger so it becomes higher than the $h$-type mode.

The dispersion curves are additionally splitted due to their coupling described by the nondiagonal matrix elements $V$ of  Hamiltonian~(\ref{eq:dispdet}).
As it is seen in Fig.~\ref{fig:Disp}, the dispersion branch for the $h$-type polaritons becomes steeper at pressure $P=0.23$~GPa. This behavior is equivalent to the mass reduction of the $h$-type polaritons.
At the same time, the dispersion curve of $l$-type becomes flatter, which  is equivalent to an increase of polariton mass. In the first approximation, this effect can be described as the convergence of effective masses of the of $h$- and $l$-types polaritons.

It should be pointed out that perturbation $V$ acts similar to a perturbation in the Hamiltonians of an electron and a hole in the topological insulators, superconducting condensate and topological crystalline insulators describing the Dirac mass generation mechanism for massless particles~\cite{Qi, Pashkin,  Liang}.

\section{Experimental reflectance spectra and their analysis}
\label{results}

We have studied a heterostructure with the $\text{GaAs/Al}_{0.3}\text{Ga}_{0.7}\text{As}$ quantum well grown by the molecular beam epitaxy technology at a [001] GaAs substrate. The QW width is $L_{QW}=240$~nm.
Besides the QW, the heterostructure contains several technological layers, which are not important for present study.

The reflectance spectra have been measured using a standard setup consisting of a white light source (an incandescent lamp), a 0.5-m monochromator, a helium closed-cycle cryostat and a photodiode.  
The sample was placed in the vacuum space of the cryostat and cooled down to temperature $T=12$~K.
Monochromatic radiation from the monochromator was directed at the small angle relative to normal to the sample surface. 
Incident light was linearly polarized along the strain [110] axis.
The reflected beam was detected by the photodiode. In the optical scheme used, the photoluminescence of the sample excited by weak monochromatic light is negligibly small and does not affect the detected signal.
A mechanical micropress was used to apply a constant stress to the sample. The magnitude of applied stress was evaluated by comparison of the spectra with theoretical simulations.

\begin{figure}[t] % figure placement: here, top, bottom, or page
\includegraphics[clip,width=.49\columnwidth]{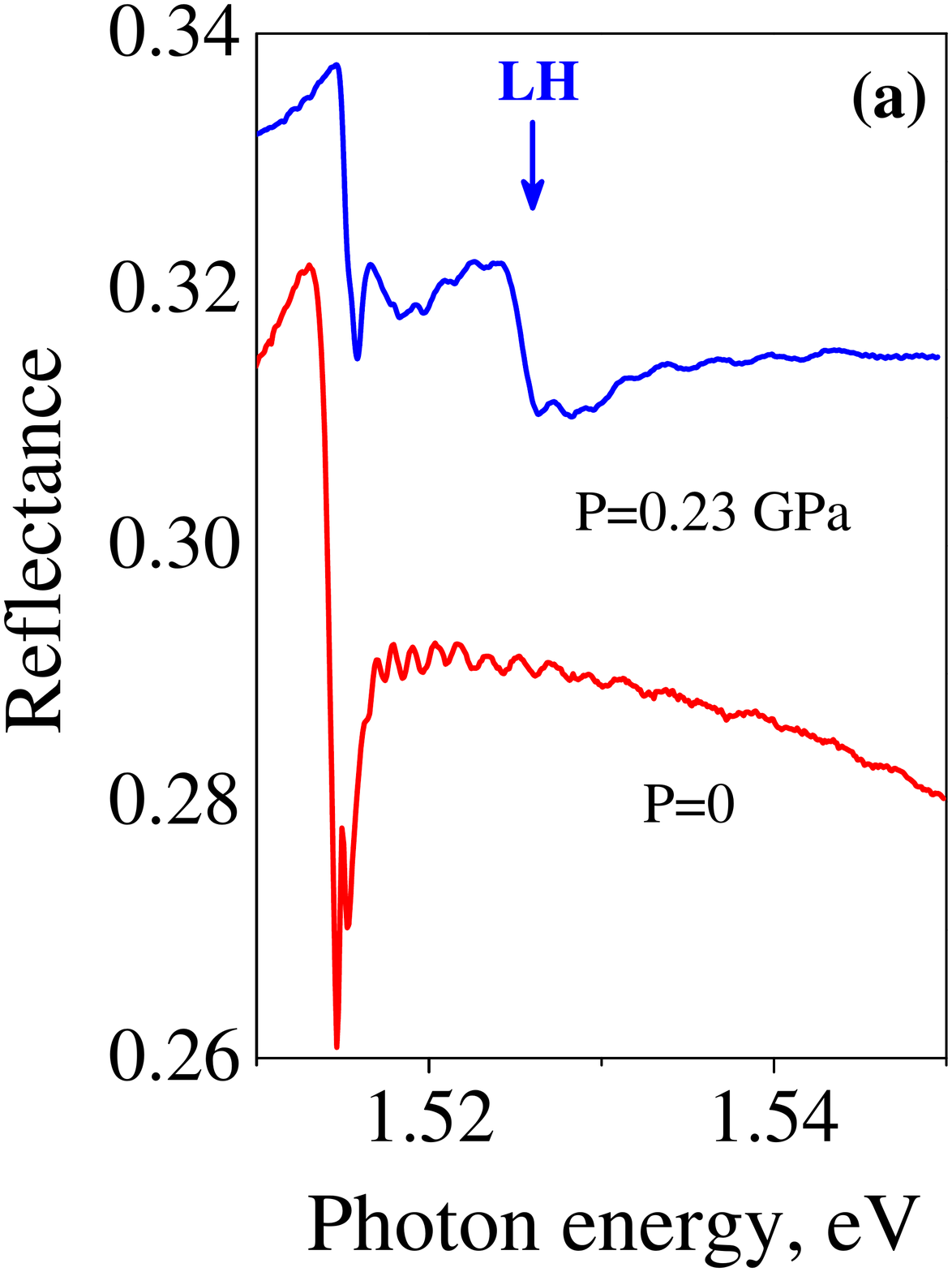}
\includegraphics[clip,width=.49\columnwidth]{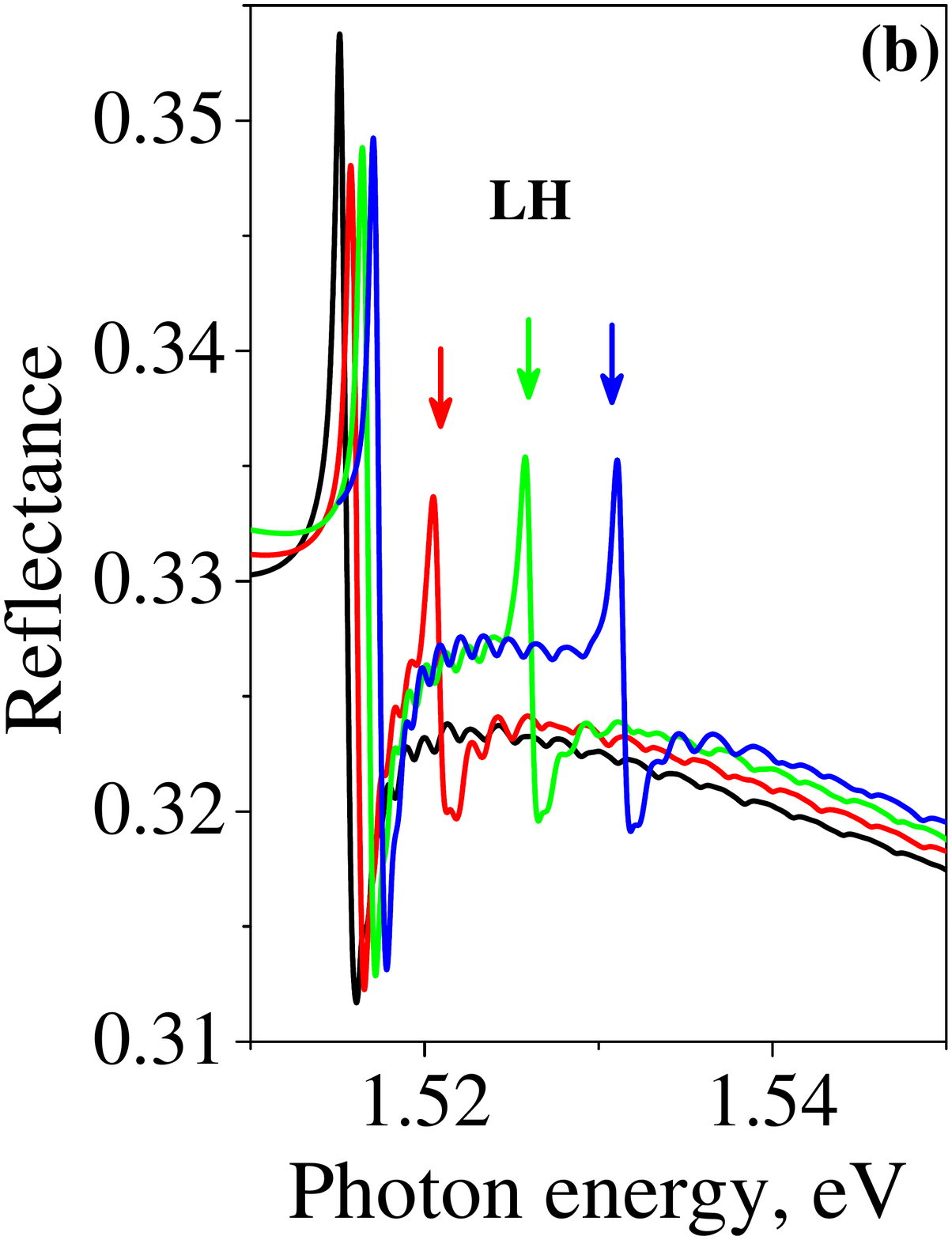}
\caption{(a) Experimentally measured reflectance spectra for the GaAs/AlGaAs QW of the width $L_{QW}=240$ nm.
Red and blue curves denote spectra measured for pressure $P=0$ and $P=0.23$~GPa, respectively.
(b) The reflectance spectra calculated for pressure $P=$ $0$, $0.1$, $0.2$ and $0.3$~GPa (black, red green  and blue, curves, respectively).
The vertical arrows indicate the spectral features related to the optical transition of the splitted off light-hole exciton (LH).}
\label{fig:exper0}
\end{figure}

Examples of measured spectra are shown in Fig.~\ref{fig:exper0}(a). 
They demonstrate spectral features caused by the interference of polaritonic waves in the QW. 
The dominant feature at the photon energy of about 1.516~eV at zero pressure is due to the interference of polaritonic waves in the range of anti-crossing of exciton-like and photon-like polaritonic modes at $\mathbf K=\mathbf q$, where $\mathbf q$ is the light wave vector.  
The quasiperiodic oscillations are caused by the interference of exciton-like and photon-like modes~\cite{LogTrifIg2014, Kavokin, Loginov}.
The effective period of these oscillations is determined by the width of the QW and by the mass of the heavy-hole exciton.

To simulate the reflectance spectra, it is necessary to consider the light wave reflected from the surface of the sample and three polariton modes propagating in the QW. 
The amplitudes of these modes can be determined if one considers the Maxwell's boundary conditions as well as the Pekar's additional boundary conditions (ABC)~\cite{Kavokin}.

The Maxwell's and Pekar's boundary conditions give rise to a system of linear equations with respect to amplitudes of the electric field of light and polaritonic waves in the structure.
The solution of this system allows one to obtain the ratio of amplitudes of the incident and reflected light waves (see, for example, Refs.~\onlinecite{LogTrifIg2014, Loginov}).
The coefficient of reflectance is the squared module of this ratio:
$$
R(\omega)=\Bigl|\frac{E_i}{E_r}\Bigr|^2.
$$

The reflectance spectra calculated for QW of the same width as in the experiment are presented in Fig.~\ref{fig:exper0}(b) for several magnitudes of applied pressure. 
In the calculations, we introduced the near-surface dead layers, $L_D= 19$~nm, for excitons in the QW according to Refs.~\onlinecite{LogUbyvovk, Schiumarini}.
The material parameters have been taken the same as for calculations of the dispersion curves shown in Fig.~\ref{fig:Disp}. We have also chosen the damping parameters to obtain the width of spectral features to be approximately equal to that observed in experiment: $\Gamma_h=0.27$ meV and $\Gamma_l=0.55$~meV.
As seen, the uniaxial stress shifts the light-hole exciton to the higher energy relative to the heavy-hole energy. The heavy-hole exciton is also slightly shifted to higher energy by the stress.

\begin{figure}[t] % figure placement: here, top, bottom, or page
\includegraphics[clip,width=1.0\columnwidth]{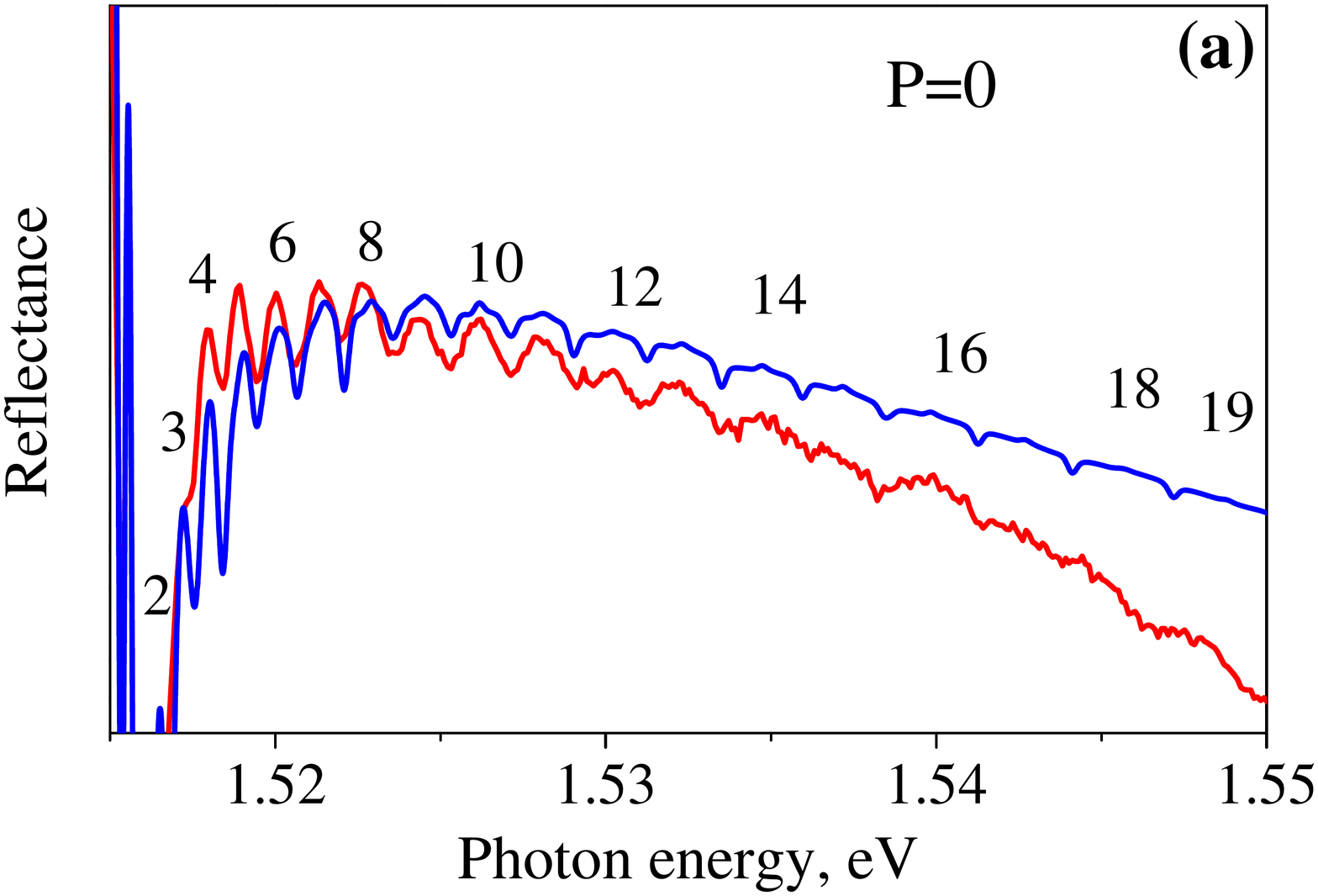}
\includegraphics[clip,width=1.0\columnwidth]{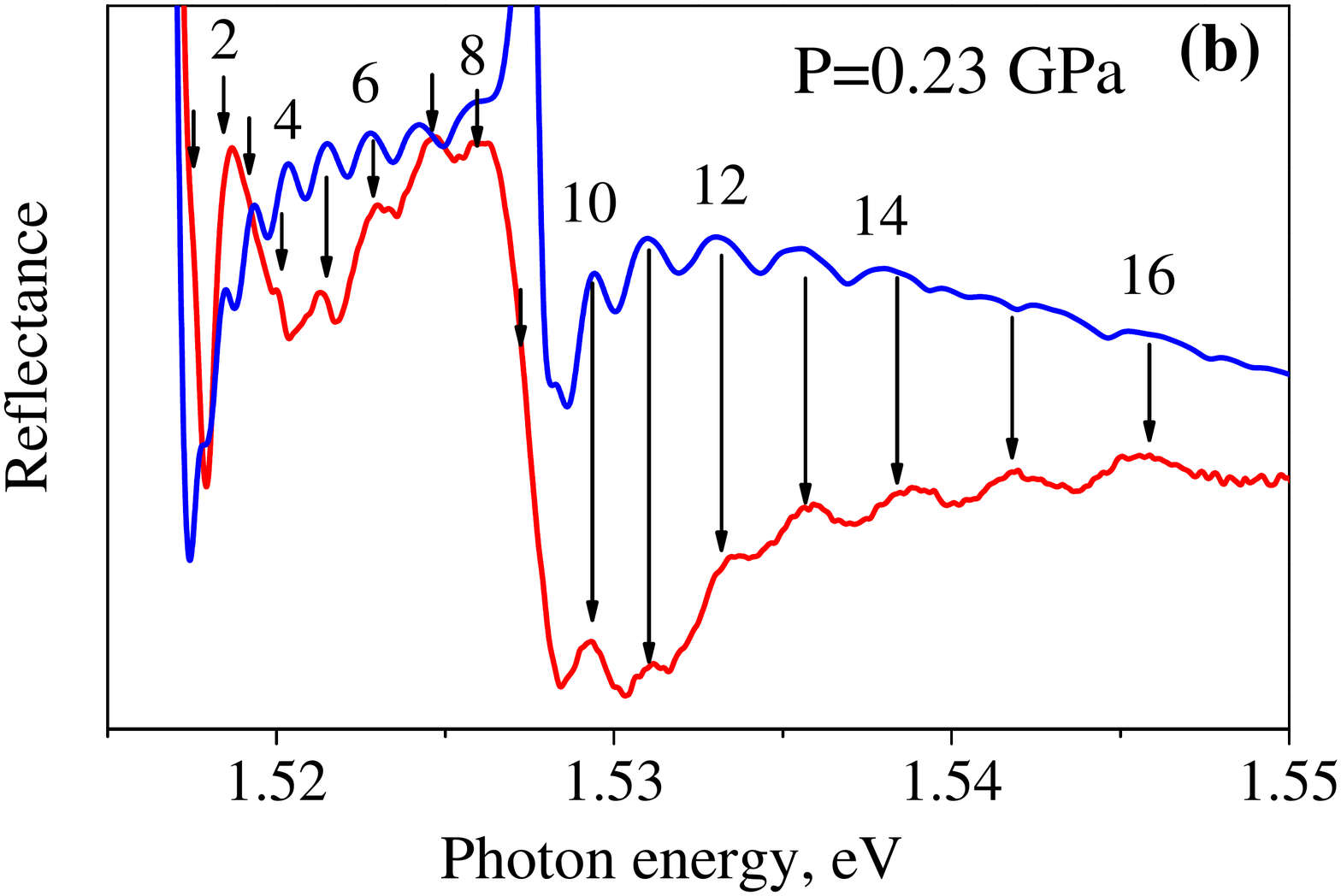}
\caption{Experimentally measured (red curves) and theoretically calculated (blue curves) reflectance spectra for the GaAs/AlGaAs QW of the width $L_{QW}=240$ nm for pressures $P=0$ (a) and $P=0.23$ GPa (b). Numbers in both panels numerate the oscillations. Vertical arrows in panel (b) indicate respective oscillations in experimental spectrum.} 
\label{fig:exper}
\end{figure}

In Fig.~\ref{fig:exper}, the calculated and experimental reflectance spectra for the energy above the ground state of the heavy-hole exciton are compared.
For more precise comparison with experiment we have used different damping parameters for the non-stressed and stressed QW:
$\Gamma_h=0.07$~meV, $\Gamma_l=0.35$~meV for $P=0$ and $\Gamma_h=0.27$~meV, $\Gamma_l=0.55$~meV for $P=0.23$~GPa. The possible reason for the increase of peak broadening with pressure is some inhomogeneity of the pressure.
It should be also noted, that the damping parameter for the light-hole exciton is larger than that for the heavy-hole one, that is typically observed in experiment  (see, Fig.~\ref{fig:exper0}(a) and Ref.~\onlinecite{Loginov}).

There is noticeable difference in general behavior of experimental and simulated spectra. We attribute this difference to the complex layer structure of the sample under study containing many technological layers. It is well known (see, e.g., Ref.~\onlinecite{Kavokin}) that the interference of light waves reflected from different layers may result in slow spectral modulation of reflectance. 
We ignore this effect in the simulation for simplicity and consider here the behavior of the spectral oscillations, which contains valuable information on the polariton dispersion.
The calculations show that the observed spectral oscillations are mainly caused by the contribution from the $h$-type modes.
The contribution of the $l$-type modes for $\mathbf K\gg\mathbf q$ is negligibly small due to the smaller oscillator strength as well as the significantly larger inhomogeneous broadening.

The calculated spectra well reproduce overall behavior of the spectral oscillations observed experimentally (see Fig.~\ref{fig:exper}). In particular, the enlargement of energy distance between the oscillations in the stressed QW is reproduced. 
The increase of the oscillation quasiperiod indicates a decrease of the effective mass $M_{h}$ of the heavy-hole exciton. This decrease at $P=0.23$ GPa is not small: $\Delta M_h=0.026$ $m_0$ which is of about 5 \% of initial exciton mass. 

\section{Conclusions}
\label{conclusion}

We have theoretically and experimentally studied the impact of uniaxial mechanic stress applied along the second order symmetry axis on the polariton reflectance spectra of the heterostructure with a wide GaAs/AlGaAs QW.
It was shown that the applied stress leads to the decrease of effective mass of the heavy-hole exciton.
In the reflectance spectra, this effect manifests itself in the increase of relative energy distance between the spectral oscillations caused by the interference of photon-like and exciton-like polariton waves in the QW.
The observed phenomenon is due to the stress-induced coupling of the heavy-hole and light-hole exciton states.
The coupling is described by the Bir-Pikus Hamiltonian and has some similarity with the Dirac mass generation mechanism for massless carriers in topological insulators and superconducting condensate.
The analysis of spectral oscillations shows a good agreement of the theory with the experiment.

\section*{Acknowledgments}

We are grateful to I. Ya. Gerlovin and M. M. Glazov for discussions. Financial support from the Russian Ministry of Science and Education (contract No. 11.G34.31.0067), SPbSU (grants No. 11.38.213.2014), and the Russian Foundation for Basic Research in the frame of International Collaborative Research Center TRR 160 is acknowledged. The authors also thank the SPbSU Resource Center ``Nanophotonics'' (www.photon.spbu.ru) for the sample studied in present work.

\end{document}